\documentclass[11pt,a4paper]{article}
\usepackage{amssymb}
\usepackage{amsmath}
\usepackage{amsfonts}
\usepackage[pdftex]{graphicx}
\graphicspath{{./graphics/}}
\DeclareGraphicsExtensions{.pdf,.png,.jpg}
\usepackage{amscd}
\usepackage{a4wide}
\usepackage{microtype}
\usepackage{bbm}
\usepackage{slashed}
\usepackage{fancyhdr}
\usepackage{amsthm}
\usepackage{mathrsfs}
\usepackage{hyperref}
\usepackage{color} 
\usepackage{pstricks}
\usepackage{epstopdf}

\definecolor{darkred}{rgb}{0.8,0.1,0.1}
\hypersetup{
     colorlinks=true,         
     linkcolor=darkred,
     citecolor=blue,
}

\usepackage{comment}
\usepackage[margin=2.7cm]{geometry}
\usepackage[all,cmtip]{xy}

\usepackage{marvosym}

\newtheorem{theorem}{\rmfamily\bfseries{Theorem}}[section]

\newtheorem{definition}[theorem]{\rmfamily\bfseries{Definition}}

\theoremstyle{remark}

\newtheorem{remark}{Remark}

\def\un{1\kern-3pt \rm I}

\newcommand{\Hilbert}{\mathscr{H}}
\newcommand{\Sd}{\mathscr{S}^\prime{}}
\newcommand{\Sdf}{\mathscr{S}{}}

\numberwithin{equation}{section}

\def\1{\mathbbm{1}}

\newcommand{\oN}{{\mathbb N}}
\newcommand{\oR}{{\mathbb R}}

\newcommand{\oC}{{\mathbb C}}

\title{\bf{On the relation between Dencker's polarization set and the gauge-fixing in electrodynamics}}

\author{Daniel H.T. Franco \vspace{4mm}\\
      Grupo de F\'\i sica-Matem\'atica e Teoria Qu\^antica dos Campos,\\
      Universidade Federal de Vi\c cosa, Departamento de F\'\i sica,\\
      Av. Peter Henry Rolfs s/n, Campus Universit\'ario,\\
      Vi\c cosa, MG, Brasil, CEP: 36570-900.\vspace{4mm}\\
      and \vspace{4mm}\\
      Centro de Estudos de F\'\i sica Te\'orica,\\
      Setor de F\'\i sica-Matem\'atica,\\
      Rua Rio Grande do Norte 1053/302, Funcion\'arios,\\
      Belo Horizonte, Minas Gerais, Brazil, CEP 30130-131.\vspace{4mm}\\
{\small e-mail: \texttt{daniel.franco@ufv.br, dhtfranco@gmail.com}} \vspace{4mm}\\
{\em Dedicated to Prof. Jos\'e Abdalla Helay\"el-Neto on the occasion of his 70th birthday.}
}

\date{\today}

\begin{document}

\maketitle

\begin{abstract}
The purpose of this short review, based in part on ideas developed in an article by the author and
Fagundes~\cite{DF}, is to emphasize that the gauge-fixing condition, necessary to eliminate the spurious 
degrees of freedom of the electromagnetic field, is elegantly handled by Dencker's work on the propagation of 
polarization sets for systems of real principal type.
\end{abstract}

\section{Introduction}
\label{Sec1}
\hspace*{\parindent}
As is well-known, in electrodynamics, the electromagnetic field can be expressed by the 4-vector
potential $A_\mu=(A_0,{\boldsymbol{A}})$, where $A_0$ is the scalar or Coulomb potential and
${\boldsymbol{A}}$ the vector potential. It is a real vector potential and, when quantized, represents
a spin-1 neutral particle, called the photon. It obeys the Maxwell equation in a vacuum
\begin{align*}
\partial_\mu F^{\mu \nu}=0\,\,,
\end{align*}
where $F^{\mu \nu}=\partial^\mu A^\nu-\partial^\nu A^\mu$ is the field strength tensor.

It has long been understood that the potential $A_\mu$ is not unique. If it is changed to
\begin{align}
A_\mu(x) \longrightarrow A'^{\,\mu}(x)=A^\mu(x)+\partial^\mu \chi(x)\,\,,
\label{potencial}
\end{align}
where the gauge function $\chi$ is any single-valued, continuously differentiable function of $x$ that vanishes at infinity,
then the field strength tensor $F^{\mu \nu}$ remains unchanged (and the Maxwell equation in a vacuum too). Therefore,
many different sets of potentials $A^\mu$ correspond to a given set of solutions of Maxwell equations in a vacuum.
In other words, the potential $A'^{\,\mu}$ is fully equivalent to the original one $A^\mu$, yielding the {\em same} electric
and magnetic fields, but satisfying different dynamical equations. To eliminate this redundancy, a gauge has to be fixed.
The gauge-fixing problem is simply the problem of choosing a representative in the class of equivalent potentials, convenient
for practical calculations or most suited to physical intuition. We are going to adopt the relativistically covariant gauge, called
{\em Lorenz} gauge
\begin{align}
\partial^\mu A_{\mu}=0
\quad \Longrightarrow \quad
\partial_t A_0+{\boldsymbol{\nabla}} \cdot {\boldsymbol{A}}=0\,\,.
\label{LorenzG}
\end{align}
It is obviously not an equation of motion to specify the space-time behavior of the vector field $A_\mu$.
There are many other gauges, but the Lorenz gauge is unique among the constraint gauges in retaining manifest
{\em Lorentz} invariance. 

\begin{remark}
The Lorenz gauge was originally named after the Danish physicist Ludvig Lorenz and not after Hendrik Lorentz.
It is often misspelled ``Lorentz gauge.'' See Ref.~\cite{Jackson}, p. 670 for the explanation.
\end{remark}

Although the potential has four components, a photon can have only two physical degrees of freedom. To drop the redundant
degrees of freedom, once adopted the Lorenz gauge the Maxwell equation becomes
\begin{align*}
\partial^\nu \partial_\nu A_{\mu}=0\,\,.
\end{align*}
Thus each component of $A^\nu$ separately obeys the Klein-Gordon equation (with $m=0$), {\em i.e.},
this choice leads to the decoupled wave equations for $A_0$ and ${\boldsymbol{A}}$. Thus, since the four components
of $A_\mu$ are not coupled in the wave equation, various other conditions are possible. For example, $A_0=0$
(temporal gauge) and ${\boldsymbol{\nabla}} \cdot {\boldsymbol{A}}=0$ (Coulomb gauge). Each of these conditions
has the following important property: if they are satisfied in an arbitrarily small time interval, they are automatically
satisfied for all $t$. This means, for example, that if $A_0=0$ and $\partial_t A_0=0$ at $t=0$ for all ${\boldsymbol{x}}$,
then $A_0$ vanishes identically for all $t$ as solution of the wave equation. Therefore, from condition (\ref{LorenzG})
it follows that ${\boldsymbol{\nabla}} \cdot {\boldsymbol{A}}=0$. This condition, written in Fourier space
\begin{align*}
{\boldsymbol{k}} \cdot {\boldsymbol{\widehat{A}}}(\boldsymbol{k})=0\,\,,
\end{align*}
expresses the fact that the electromagnetic field is {\em transversal} (hence there are only 2 degrees of freedom, the
two transverse {\em polarization vectors} at each ${\boldsymbol{k}}$ point). There are no longitudinal and scalar photons in
this gauge.

The best way to specify the transverse property of the electromagnetic field is by means of polarization vectors.
Writing the polarization vectors of the photon having energy-momentum $k$ as $\varepsilon_\mu(k)$,
the potential $A_\mu$ can be expanded in Fourier space as
\begin{align}
\widehat{A}_\mu(k)=\sum_\lambda \frac{\varepsilon_\mu(k,\lambda)}{\sqrt{2\omega V}}
\Bigl(a_\lambda(k) e^{i\,kx}+a^*_\lambda(k) e^{-i\,kx}\Bigr)
\label{FourierExp}
\end{align}
Of these four polarization vectors, we pick $\varepsilon_0$ to be time-like, while $\varepsilon_j$, 
with $j=1,2,3$, are space-like. Hence, the Lorenz condition (\ref{LorenzG}) implies
\begin{align*}
k^\mu \varepsilon_\mu(k,\lambda)=0\,\,.
\end{align*}

Equation (\ref{potencial}) means $\varepsilon_\mu^{~\nu}(k)$ has an extra degree of freedom to change
\begin{align*}
\varepsilon_\mu(k,\lambda) \longrightarrow \varepsilon'_\mu(k,\lambda)
=\varepsilon_\mu(k,\lambda)+i k_\mu \widehat{\chi}(k)\,\,,
\quad k^2=k_\mu k^\mu=0 \quad \text{or} \quad \omega=k_0=|{\boldsymbol{k}}|\,\,.
\end{align*}
We can choose $\widehat{\chi}(k)$ to make $\varepsilon_0=0$. Then the Lorenz condition becomes
\begin{align}
{\boldsymbol{k}} \cdot {\boldsymbol{\varepsilon}}(\boldsymbol{k})=0\,\,.
\label{CoulombG}
\end{align}
Hence the free electromagnetic field ({\em i.e.} the electromagnetic wave) has only transverse po\-la\-ri\-zation, with two
degrees of freedom as mentioned. In other words, the solutions to free electromagnetic field equations are given by the
polarization in which the solutions are {\em strongly oscillating}.

In order to prove (\ref{CoulombG}), we will choose a slightly different (but equivalent) point of view
based on Dencker's work on the propagation of polarization sets for systems of real principal type~\cite{Dencker},
which refine the notion of wave front set for vector-valued distributions.\footnote{The term ``wave front set'' was coined
by Lars H\"ormander around 1970~\cite{Hor} (see also~\cite{DH,Hor1}).} We will show that the gauge-fixing condition,
necessary to eliminate the spurious degrees of freedom of the electromagnetic field, is elegantly handled by Dencker's
work~\cite{Dencker}. We emphasize that our construction is not meant as ``new physics.'' By using directly the insights
and arguments found in~\cite{DF}, our intention is only to provide a means of proving the connection between Dencker's
polarization set and the operation of gauge-fixing in electrodynamics in manifestly-covariant formulation using potentials in
Lorenz gauge (Dencker addressed the same problem in non-covariant formalism). To be specific, we shall show that the
components of the potential $A_\mu$ in which the solutions to Maxwell equations are {\em strongly oscillating} (namely the
directions of the strongest singularities) are directly related to the physical transverse polarization states.

We dedicate this article to Prof. Jos\'e Abdalla Helay\"el-Neto, a physicist of extraordinary perception and influence, and an
expert in electromagnetism.

\section{\bf Propagation of polarizations in electrodynamics}
\label{Sec4}
\hspace*{\parindent}
In this section, we will make use of the polarization set which was introduced by Dencker~\cite{Dencker} in
order to analyze the singularities of vector-valued distributions. Applying Dencker's results, we will show that
the singularities of solutions of the Maxwell equations propagate in a simple way.

To begin with, we first give the definition of Hilbert space valued 
distributions.\footnote{The presentation of the material is based on 
Ref.~\cite{Schwartz}.} Assume we are given a Hilbert space $\Hilbert$. 

\begin{definition}
The space ${\Sd}(\oR^n,\Hilbert)$ of $\Hilbert$-valued distributions,
or of distributions with values in $\Hilbert$, is defined 
to be the set of all weakly continuous linear maps
$\Sdf(\oR^n) \to \Hilbert$.
\end{definition}

Note that due to the nuclearity of $\Sdf(\oR^n)$ these maps are automatically
strongly continuous. If $X \subseteq \oR^n$ is an open subset, a linear map
$u:\Sdf(X) \to \Hilbert$ is in $\Sd(X,\Hilbert)$ if, and only if, there is a constant 
$C>0$ such that
\[
\bigl|u(\varphi)\bigr| \leq C \sup_{x \in X} \Bigl\{(1+|x|)^\alpha
\bigl|D_x^\beta \varphi(x)\bigr|\Bigr\} < \infty\,\,,\quad \alpha,\beta \in \oN^n_0\,\,, 
\]
for all $\varphi \in \Sdf(X)$.

In order to define the polarization set of a vector-valued distribution
we take in the definition of Hilbert space valued distributions
$\Hilbert \equiv \oC^N$. Let $u=(u_1,\ldots,u_N) \in {\mathscr S}^\prime(\oR^n,\oC^N)$ 
an $N$-dimensional vector-valued tempered distribution, {\it i.e.}, a vector of scalar
distributions $u_j \in {\mathscr S}^\prime(\oR^n)$, $j=1,\ldots,N$.
We should note that the wave front set of a vector-valued distribution 
$u=(u_1,\ldots,u_N) \in {\mathscr S}^\prime(\oR^n,\oC^N)$ is just defined as the union 
of the wave front sets of all its components, $WF(u)=\cup_j WF(f_j)$, but this gives no 
information about which components of $u$ that are singular.

\begin{definition}[Definition 2.1 in~\cite{Dencker}]
The polarization set of a distribution $u \in {\mathscr S}^\prime(\oR^n,\oC^N)$
is defined by
\begin{align*}
WF^{\rm pol}(u)=\bigcap_{Au \in C^\infty}\Bigl\{(x,k;\omega(x,k))
\in (T^*(\oR^n) \setminus 0) \times \oC^N \mid \omega(x,k) \in 
{\rm Ker}\,\sigma(A)(x,k)\Bigr\}\,\,,
\end{align*}
where $A$ is a $1 \times N$ system of classical pseudodifferential operators
with principal symbol $\sigma(A)$.
\label{defPol}
\end{definition}

For a scalar distribution ($N=1$) the polarization set contains the same information 
as the wave front set. For arbitrary $N$, the relation between the polarization set 
and the wave front set of a vector-valued distribution is obtained by
projecting the nontrivial points onto the cotangent bundle.

\begin{theorem}[Proposition 2.5 in~\cite{Dencker}]
Let $u \in {\mathscr S}^\prime(\oR^n,\oC^N)$ and $\pi: (T^*(\oR^n) \setminus 0) 
\times \oC^N \to (T^*(\oR^n) \setminus 0)$ be the projection onto the cotangent bundle: 
$\pi:(x,k;\omega(x,k))=(x,k)$. Then
\begin{equation*}
\pi \Bigl(WF^{\rm pol}(u) \setminus \bigl((T^*(\oR^n) \setminus 0) 
\times 0\bigr)\Bigr)=WF(u)\,\,.
\end{equation*}
\end{theorem}

The above theorem clearly shows that the polarization set is a refinement of the wave front set. 
In addition, it contains information about the directions in the additional vector 
space in which the distribution is singular.

\begin{definition}
A pseudodifferential operator $P$ on $\oR^n$ is said to be of {\bf real
principal type} if its principal symbol $p(x,k)$ is real and for $p=0$ the 
Hamiltonian vector field $H_p$,
\begin{equation*} 
H_p(x,k)=\sum_\mu \left(\frac{\partial
p(x,k)}{\partial k_\mu}\frac{\partial}{\partial x^\mu} -
\frac{\partial p(x,k)}{\partial x^\mu}\frac{\partial}{\partial k_\mu}\right)\,\,,
\end{equation*}
does not vanish nor does it have the radial direction, that is
$H_p \neq -\frac{\partial p}{\partial x^\mu}\frac{\partial}{\partial k_\mu}$.
\end{definition}

\begin{remark}
The principal symbols are homogeneous since we only consider
classical pseudodifferential operators. Thus, the polarization set is closed,
conical in the $k$ variables, and linear in the fiber.
\end{remark}

An important result shown in~\cite{Dencker} is the propagation of polarization sets 
for systems of real principal type.

\begin{definition}
An $N \times N$ system $P$ of pseudodifferential operators on $\oR^n$ with principal 
symbol $p(x,k)$ is of real principal type at $(x_0,k_0) \in T^*(\oR^n) \setminus 0$ if
there exists an $N \times N$ symbol $\widetilde{p}(x,k)$ such that
\begin{equation*}
\widetilde{p}(x,k)p(x,k)=q(x,k) \cdot \un\,\,,
\end{equation*}
in a neighborhood of $(x_0,k_0)$, where $q(x,k)$ is a scalar symbol of real principal
type and $\un$ is the identity in $\oC^N$. 
\end{definition}

Let ${\mathscr N}_p=\bigl\{(x,k;\omega(x,k)) \in (T^*(\oR^n) \setminus 0) \times 
\oC^N \mid \omega(x,k) \in {\rm Ker}\,p(x,k)\bigr\}$, then the {\it characteristic} set 
of a system of pseudodifferential operators, $P$, on $\oR^n$ with principal symbol $p(x,k)$ 
is defined as
\[
{\rm Char}(P)=\bigl\{(x,k) \in T^*(\oR^n) \setminus 0 \mid {\rm det}\,p(x,k)=0\bigr\}=
\pi \Bigl({\mathscr N}_p \setminus \bigl\{(x,k;0)\bigr\}\Bigr)\,\,.
\]
According to Definition \ref{defPol}, if $Pu \in C^\infty$, then we must have 
$WF^{\rm pol}(u) \subseteq {\mathscr N}_p$. This implies that polarization sets of 
$u$ satisfying $Pu \in C^\infty$ can only propagate in 
${\mathscr N}_p$.\footnote{For a scalar distribution the characteristic set 
of a pseudodifferential operators $P$ on $\oR^n$, with principal symbol 
$p(x,k)$, is defined as
\[
{\rm Char}(P)=\bigl\{(x,k) \in T^*(\oR^n) \setminus 0 \mid p(x,k)=0\bigr\}\,\,,
\]
and can be interpreted as the set of all directions $k$ suppressed by $P$ to leading
order, at a point $x$. According to Duistermaat and H\"ormander~\cite{DH}, 
if $P$ is a pseudodifferential operators on $\oR^n$ with principal symbol 
$p(x,k)$ and $f \in {\mathscr S}^\prime(\oR^n)$ such that $Pf \in C^\infty$, then
$WF(f) \subset {\rm Char}(P)$. If furthermore $P$ is of real principal type, then 
$WF(f)$ is invariant under the flow generated by the Hamiltonian vector field of $p$.}
To state the propagation of polarization sets for systems of the real principal type more 
precisely, Dencker introduced the notion of Hamilton orbits.

\begin{definition}
Let $\gamma$ be a null-bicharacteristic curve in ${\rm Char}(P)$. A Hamilton orbit of a
system $P$ of real principal type is defined as $\bigl\{(x,k;z\omega(x,k)) \mid \omega(x,k) 
\in \gamma, z \in \oC \bigr\} \subseteq {\mathscr N}_p$, where $\omega(x,k)$ is a $C^\infty$-function 
satisfying $D_p \omega(x,k)=0$, with the {\bf Dencker's connection} $D_p \omega(x,k)$ given by
\begin{equation*}
D_p \omega(x,k)=H_q\omega(x,k)+\frac{1}{2}\{\widetilde{p},p\}\omega(x,k)+
i\,\widetilde{p}p^{\rm s}\omega(x,k)\,\,,
\end{equation*} 
where 
\begin{equation*}
H_q(x,k)=\sum_\mu \left(\frac{\partial
q(x,k)}{\partial k_\mu}\frac{\partial}{\partial x^\mu} -
\frac{\partial q(x,k)}{\partial x^\mu}\frac{\partial}{\partial k_\mu}\right)\,\,,
\end{equation*}
is the Hamilton field associated with $q$, 
\begin{equation*} 
\{\widetilde{p},p\}= \sum_\mu\frac{\partial
\widetilde{p}(x,k)}{\partial k_\mu}\frac{\partial p(x,k)}{\partial x^\mu}
-\sum_\mu\frac{\partial \widetilde{p}(x,k)}{\partial{x^\mu}}\frac{\partial
p(x,k)}{\partial {k_\mu}}\,\,,
\end{equation*}
is the Poisson bracket, and 
\begin{equation*}  
p^{\rm s}(x,k)=p_{m-1}(x,k)-\frac{1}{2i}\sum_\mu\frac{\partial^2
p(x,k)}{\partial x^\mu \partial k_\mu}\,\,, 
\end{equation*}
is the subprincipal symbol of $P$.\footnote{Recall that the symbol of $P$ is 
a sum of homogeneous terms,
\[ 
\sigma(P)(x,k)=p(x,k)+p_{m-1}(x,k)+p_{m-2}(x,k)+\cdots\,\,, 
\]
where $p(x,k)=\sigma_P(P)$ is the principal symbol, and $p_j(x,k)$ is
homogeneous of order $j$.}
\end{definition}

\begin{remark}
It should be emphasized that the Hamilton orbits are independent of the choice of 
$\widetilde{p}$. Moreover, the polarization vectors over the points in the wavefront 
set follow a simple parallel transport law along the bicharacteristics that form the 
wave front set. This means that for every vector field $\omega$ along a bicharacteristic
$\gamma$ in ${\rm Char}(P)$ we have $D_p \omega \in {\rm Ker}\,p$ along $\gamma$ if and
only if $\omega \in {\rm Ker}\,p$ along $\gamma$. The equation $D_p \omega=0$ can then be
solved with $(x,k;\omega) \in {\mathscr N}_p$, which is a necessary
condition for elements of the polarization set. See~\cite{Dencker} for details!
\end{remark}

Having defined Hamilton orbits, the propagation of polarization sets for systems of
principal type is stated as follows:

\begin{theorem}[Theorem on the propagation of singularities for vector-valued 
distributions]
Let $P$ be a system of real principal type on $\oR^n$. If $Pu \in C^\infty$, then
$WF^{\rm pol}(u)$ is a union of Hamilton orbits of $P$.
\label{theoP}
\end{theorem}

In addition, the Theorem \ref{theoP} gives us that the set $\pi \Bigl(WF^{\rm pol}(u) 
\setminus \bigl((T^*(\oR^n) \setminus 0) \times 0\bigr)\Bigr)=WF(u)$ is invariant under 
the shifts along the trajectories of the Hamiltonian system
\begin{align*}
\begin{cases}
\dot{k}=-\displaystyle{\frac{\partial p(x,k)}{\partial x}}\,\,, \\[3mm]
\dot{x}=\displaystyle{\frac{\partial p(x,k)}{\partial k}}\,\,.
\end{cases}
\end{align*}

Any theory of electrodynamics determines the motion of electromagnetic waves
through the corresponding system of partial differential equations. 
Electromagnetic wave trajectories can be obtained in the geometric optical limit 
by studying the propagation of {\bf singularities} of the electromagnetic fields.
In the following, we would like to apply Theorem \ref{theoP} to the Maxwell equations 
in {\em free} space. But first, we must introduce the tool that implements the extension 
of our analysis to the vector-valued distributions. 

\begin{definition}
Let $q_0=(x_0,k_0)$. For $u \in {\mathscr S}^\prime(\oR^n,\oC^N)$, we say that 
$(q_0,z_0) \notin WF^{\rm pol}(u) \subseteq (T^*(\oR^n) \setminus 0) \times \oC^N$ if
there exists a pseudodifferential operator $P$ with {\it constant} coefficient on $\oR^n$  
and principal symbol $p(k)$, such that $p(k_0)z_0 \not= 0$.
\end{definition}

We will now investigate the Maxwell equations on a space-time $(\oR^{1,3},\eta)$,
where $\eta$ is the flat metric. We start considering the equation $\partial^\nu \partial_\nu A_\mu=0$.
In the Fourier space this equation takes the following form:
\begin{align*}
k^2 \widehat{A}_\mu(k)=\eta^{\alpha\beta}k_\alpha k_\beta \delta_\mu^{~\nu}\widehat{A}_\nu(k)=0\,\,.
\end{align*}
For a nontrivial solution this implies that $k^2=k_0^2-|\boldsymbol{k}|^2=0$, where 
$k^\mu=(k^0,-\boldsymbol{k})$ is a vector light-like. Note that the principal symbol of $P$ is
\begin{align*}
p(k)=\eta^{\alpha\beta}k_\alpha k_\beta \delta_\mu^{~\nu}\,\,.
\end{align*} 
so that we can choose
\[
\bigl(\widetilde{p}(k)\bigr)_\nu^{~\mu}=\delta_\nu^{~\mu}\,\,,
\quad q(k)=\eta^{\alpha\beta}k_\alpha k_\beta\,\,,
\]
in order to obtain $\widetilde{p} p=q \cdot \un$. In this way, we see that $q$ is a scalar symbol 
of real principal type; therefore $P$ is of real principal type. Also, it is easy to 
see that
\begin{align*}
{\rm Char}(P)=\bigl\{(t,\boldsymbol{x},k_0,\boldsymbol{k}) \in (T^*(\oR) \setminus 0)
\times (T^*(\oR^3) \setminus 0) \mid q(k)=0\bigr\}\,\,.
\end{align*}
This implies that $\eta^{\alpha\beta}k_\alpha k_\beta=0$, which is the surface of the light-cone.
Moreover, for $k^2=0$, $q(k)=\eta^{\alpha \beta} k_\alpha k_\beta =k^2$ is a Hamiltonian 
which generates null geodesics. To see this, consider Hamilton's equations
\[
\frac{dx^\mu}{d\tau}=\frac{\partial q}{\partial k_\mu}
\quad{\mbox{and}}\quad
\frac{dk_\nu}{d\tau}=-\frac{\partial q}{\partial x^\nu}\,\,.
\]
So, it follows that
\[
\frac{dx^\mu}{d\tau}=2\eta^{\mu\nu}k_\nu
\quad{\mbox{and}}\quad
\frac{dk_\nu}{d\tau}=0\,\,.
\]
Note that the condition $q(k)=0$ imposes that $\frac{1}{4}\eta_{\mu\nu}\dot{x}^\mu(\tau) \dot{x}^\nu(\tau)=0$
for all $\tau$ in the domain of the curve. Hence, the curve $\tau \mapsto x(\tau)$ is a null curve.
To see that this curve is a null geodesic, we recall that on a Minkowskian's space-time
the geodesic equation is exactly the straight equation, solution of the curve 
$\ddot{x}(\tau)=0$. Moreover, using the fact that
\[
k_\nu=\frac{1}{2}\eta_{\nu\mu}\frac{dx^\mu}{d\tau}\,\,,
\]
we obtain that
\[
\frac{dk_\nu}{d\tau}=\frac{d}{d\tau}\left(\frac{dx^\mu}{d\tau}\right)=0\,\,,
\]
which is the geodesic equation.

In the coordinate system where the $x_3$-axis is taken along the photon momentum $\boldsymbol{k}$,
we can take the following basis in (\ref{FourierExp}):
\begin{align}
\varepsilon_\mu(k,0)=
\left(
\begin{array}{c}
1 \\ 0 \\ 0 \\ 0
\end{array}
\right)\,\,,\,\,
\varepsilon_\mu(k,1)=
\left(
\begin{array}{c}
0 \\ 1 \\ 0 \\ 0
\end{array}
\right)\,\,,\,\,
\varepsilon_\mu(k,2)=
\left(
\begin{array}{c}
0 \\ 0 \\ 1 \\ 0
\end{array}
\right)\,\,,\,\,
\varepsilon_\mu(k,3)=
\left(
\begin{array}{c}
0 \\ 0 \\ 0 \\ 1
\end{array}
\right)\,\,.
\end{align}
This basis of polarization vectors form a four-dimensional orthonormal system
satisfying the relation
\[
\varepsilon_\mu(k,\lambda) \varepsilon^\mu(k,\lambda')=\eta_{\lambda \lambda'}\,\,. 
\]
Then the completeness condition or sum of the polarization vectors is written as
\begin{align}
\sum_{\lambda,\lambda'} \varepsilon_\mu(k,\lambda) \varepsilon_\nu(k,\lambda')
\eta_{\lambda \lambda'}=\eta_{\mu \nu}\,\,.
\label{epsiloneta} 
\end{align}

At this point, we wish to find the elements of the set ${\rm Char}(P)$,
with $P=\partial_\mu \partial^\mu$, such that $q(k)=0$. For this, we replace (\ref{epsiloneta}) in 
the expression of $q(k)$. Immediately, we obtain
\begin{align}
q(k)=\Bigl(k^\mu \varepsilon_\mu(k,\lambda) \Bigr)
\Bigl(k^\nu \varepsilon_\nu(k,\lambda') \Bigr)
\eta_{\lambda \lambda'}=0\,\,.
\label{Cha(P)}
\end{align}
In order to investigate the consequences of this, we shall consider for simplicity $\boldsymbol{k}$
along the $x_3$-direction. In this case $k^\mu=(|\boldsymbol{k}|;0,0,\boldsymbol{k})$. Hence, from
expression (\ref{Cha(P)}) it follows immediately that the space-like polarizations, $\varepsilon_\mu(k,1)$ and
$\varepsilon_\mu(k,2)$ (the physical transverse polarization states) satisfy the condition
\begin{align}
k^\mu \varepsilon_\mu(k,1)=k^\mu \varepsilon_\mu(k,2)=0\,\,.
\label{kernel3}
\end{align}
This establishes that
\begin{align}
{\rm Ker}\,p(k)=\Bigl\{\varepsilon_\mu(k,\lambda) \in \oC^4 \mid 
k^\mu \varepsilon_\mu(k,\lambda)=0, \lambda=1,2 \Bigr\}\,\,.
\label{kernel1}
\end{align}
Therefore, the vector $\varepsilon_\mu(k,3)$ (the longitudinal polarization state) and the vector $\varepsilon_\mu(k,0)$
(the time-like polarization state) {\bf do not belong} to the ${\rm Ker}\,p(k)$. This shows that the components of $A_\mu$
in which the solutions to Maxwell equations are {\em strongly oscillating} (namely the directions of the strongest singularities)
are directly related to the physical transverse polarization states. Indeed, as seen in the Introduction,
the above result can be seen as a consequence of the gauge symmetry of the Maxwell equations: we can by
a suitable choice change the polarization vectors in order to gauge away the longitudinal polarization state and 
the time-like polarization state; but we can not change the transverse parts. Thus, all physical information is contained
in the transverse parts $\varepsilon_\mu(k,1)$ and $\varepsilon_\mu(k,2)$! This result demonstrates the connection
between the Dencker's polarization set and the gauge-fixing in electrodynamics.

\begin{remark}
We could have chosen to take as a solution of Eq.(\ref{kernel3}) the right- and 
left-handed polarization vectors $\varepsilon_\mu(k,\pm)=(0;1,\pm\,i,0)/\sqrt{2}$
instead of the linear polarization vectors $\varepsilon_\mu(k,1)$ and $\varepsilon_\mu(k,2)$.
\end{remark}

\begin{remark}[{\bf Propagation of light in a straight line in a homogeneous space}]
Let ${\mathscr N}_p=\bigl\{(t,\boldsymbol{x},k_0,\boldsymbol{k};\omega(t,\boldsymbol{x},k_0,\boldsymbol{k})) 
\in (T^*(\oR) \setminus 0) \times (T^*(\oR^3) \setminus 0) \times \oC^4 
\mid \omega(t,\boldsymbol{x},k_0,\boldsymbol{k}) \in {\rm Ker}\,p(k)\bigr\}$. Then, according to
Theorem \ref{theoP}, the Hamilton orbits $\gamma$ (where $\gamma$ are the $q$ 
null-bicharacteristics) associated with ${\mathscr N}_p$ are such that $D_p \omega=0$,
which means that $\omega$ is parallel transported along the bicharacteristics
$\gamma$ that form the wave front set. Through simple computation we can see that
\begin{align}
D_p \omega=2 \eta^{\mu\nu} k_\mu \cdot
\frac{\partial \omega}{\partial x^\nu}\,\,.
\label{kernel2}
\end{align}
Using Hamilton's equations, then
\[
D_p \omega=\frac{d\omega}{d\tau}\,\,.
\]

Thus, following the theorem on the propagation of singularities for vector-valued 
distributions (Theorem \ref{theoP}), the polarization set of a solution of Maxwell equations 
(in flat space-time) is a union of Hamilton orbits, {\it i.e.}, it consists of
curves $(x,k;\omega)(\tau) \in (T^*(\oR) \setminus 0) \times (T^*(\oR^3) \setminus 0) 
\times \oC^4$, parameterized by $\tau$, such that $x(\tau)$ is a straight line 
(the union of all straight lines $x(\tau)$, which can be regarded passing through 
the origin, forming the surface of the light-cone given by $x^2(\tau)=0$), $k(\tau)$ 
is tangent to the straight line $x(\tau)$ (or tangent to the surface of the light-cone
$x^2(\tau)=0$) and $\omega(\tau)$ gives the physical transverse polarization 
states (being parallel transported along the line $x(\tau)$, or along the light-cone 
$x^2(\tau)=0$). As predicted by electrodynamics, this result can be interpreted as the
well-known fact that light propagates in a straight line in a homogeneous space.
\end{remark}

\section*{Data Availability}
\hspace*{\parindent}
The data that support the findings of this study are available from the corresponding author upon reasonable request.




\begin{thebibliography}{99}
\bibitem{Jackson} D. Jackson and L. B. Okun, ``{\em Historical roots of gauge invariance,}'' {\bf Rev. Mod. Phys. 73} (2001) 663.

\bibitem{Dencker} N. Dencker, ``{\em On the propagation of polarization sets for systems of real principal type,}'' 
{\bf J. Funct. Anal. 46} (1982) 351.

\bibitem{DF} D.H.T. Franco and F.N. Fagundes, ``{\em A note on directional wavelet transform and the propagation of polarization
sets for solutions of Maxwell's equations,}'' {\bf Rev. Math. Phys. 26} (2014) 1430004. 

\bibitem{Hor} L. H\"ormander, ``{\em Fourier integral operators I,}'' 
{\bf Acta Math. 127} (1971) 79.

\bibitem{DH} J. J. Duistermaat and L. H\"ormander, ``{\em Fourier integral operators II,}'' 
{\bf Acta Math. 128} (1972) 183.

\bibitem{Hor1} L. H\"ormander, ``{\em The analysis of linear partial differential equations I:
distribution theory and Fourier analysis,}'' Springer, 1990.

\bibitem{Schwartz} L. Schwartz, ``{\em Lectures on partial differential equations and representations of
semi-groups,}'' Notes by K. Varadarajan, Tata Institute of Fundamental Research, Bombay, 1957.

\end{thebibliography}
\end{document}